\documentstyle[11pt,newpasp,twoside,epsf]{article}
\def\edcomment#1{\iffalse\marginpar{\raggedright\sl#1\/}\else\relax\fi}
\reversemarginpar

\begin{document}
\title{Triggers and alerts with GLAST}
\author{J. Cohen-Tanugi}
\affil{INFN-Pisa, via Livornese 1291, 56010 San Piero A Grado (PI)}
\author{N. Omodei}
\affil{INFN-Pisa and Universita di Siena}
\author{F. Longo}
\affil{INFN-Trieste and Dipartimento di Fisica, Universita di Trieste }
\author{S. Bansal, J. Bonnell, J.P. Norris}
\affil{Laboratory of High Energy Astrophysics, Goddard Space Flight Center, Greenbelt, MD 20771}
\author{J.D. Scargle}
\affil{NASA Ames Research Center, Moffet Field, CA  94035-1000}
\author{R. Preece}
\affil{Department of Physics, University of Alabama in Huntsville, 320 Sparkman Drive, Huntsville, AL  35805}
\author{R.M. Kippen}
\affil{NIS-2, Mail Stop D436, Los Alamos National Laboratory, Los Alamos, NM 87545}

\begin{abstract}
%We present preliminary results on Gamma Ray Burst (GRB) triggers with GLAST. 
%After a brief summary of the detector layout, GLAST expected performances on GRB 
%detection are recalled. Status report on the simulation software and preliminary 
%triggers studies are then reported, concluding that approximately 66\% 
%of GRB events will be reconstructed with less than a second delay, and
%with fewer than one false trigger every 3 day.
We present preliminary results on Gamma Ray Burst (GRB) triggers with the 
Gamma-ray Large Area Space Telescope (GLAST). 
After a brief summary of the detector layout, GLAST expected performances on 
GRB detection are recalled. 
Status report on the simulation software and preliminary 
triggers studies are then reported, already showing significant improvement on
EGRET results.
\end{abstract}

\section{Description of the instruments}
GLAST is a next generation high-energy gamma ray observatory, onboard a 
satellite scheduled for launch in 2006. It consists of a Large Area Telescope (LAT, see Brez, S. et al. 2001) and a GLAST Burst Monitor (GBM, see 
Kippen et al. 2001),
and is designed for making observations of gamma-ray sources in the 10\ keV to 300\ GeV energy range. 
%The GLAST collaboration gathers about one hundred people from 28 countries.
%It is a follow on the CGRO experiment, operational between 1991 and 1999.
%Italy is responsible for the construction of the tracker, 
%the development of part of the simulation and analysis software,
%and provides members to the Senior Scientis Advisory Comittee.

The LAT is a pair conversion telescope, operating from 10 MeV to more than 300 GeV. It is composed of three subsystems: 
the Anti-Coincidence Detector (ACD) is responsible for vetoing the enormous charged cosmic-ray background and the Earth 
albedo secondary electrons and nuclei. 
%It is composed of segmented plastic scintillator tiles, read out by wave-shifting fibers and photomultiplier tubes.
The tracker (TKR) consists of a four-by-four array of tower modules, each of which being composed of 19 pairs of interleaved 
planes of silicon strip detectors and tungsten converter sheets. The silicon strip detectors track the electron positron 
pair created by the incident gamma.
The calorimeter (CAL) is a segmented arrangement of CsI(Tl) bars, designed to give both longitudinal and transverse 
information on the energy deposition pattern.

The tracker will have the ability to determine the location of an object in the sky to within 0.5 to 5 arc minutes. 
The pair conversion signature is also used to help reject the background of charged cosmic rays.
The GBM provides overlapping energy coverage in the range 10 keV to 25 MeV for bright transients such as bursts and solar flares.
It is composed of 4 triplets of NaI(Tl) scintillators(low energy band) and 2 BGO scintillators (high energy band).

Due to its angular resolution (5 times better than EGRET) and its improved afterglow sensitivity, GLAST will extend the 
knowledge about GRB's. Its large field of view (about 2 sr) and its improved effective area (10 times EGRET's) will provide better 
burst statistics (up to 100 bursts/year with many more photons detected.)

The GBM will enhance GRB science with GLAST by providing low energy context measurements with high time resolution. This will improve 
GLAST's wide band spectral sensitivity, allow for comparison of low energy versus high energy temporal variability, and provide a 
continuity with current GRB knowledge base (e.g GRO-BATSE.)
Moreover, the GBM provides fast time/location triggers: within 2 seconds, it will provide approximate locations for ground 
or space follow up observations. In case of specifically interesting events, it will allow, within ten minutes, for repointing 
of GLAST for afterglow observations.

\section{GLAST and GRB Simulations}

\begin{figure}
\plottwo{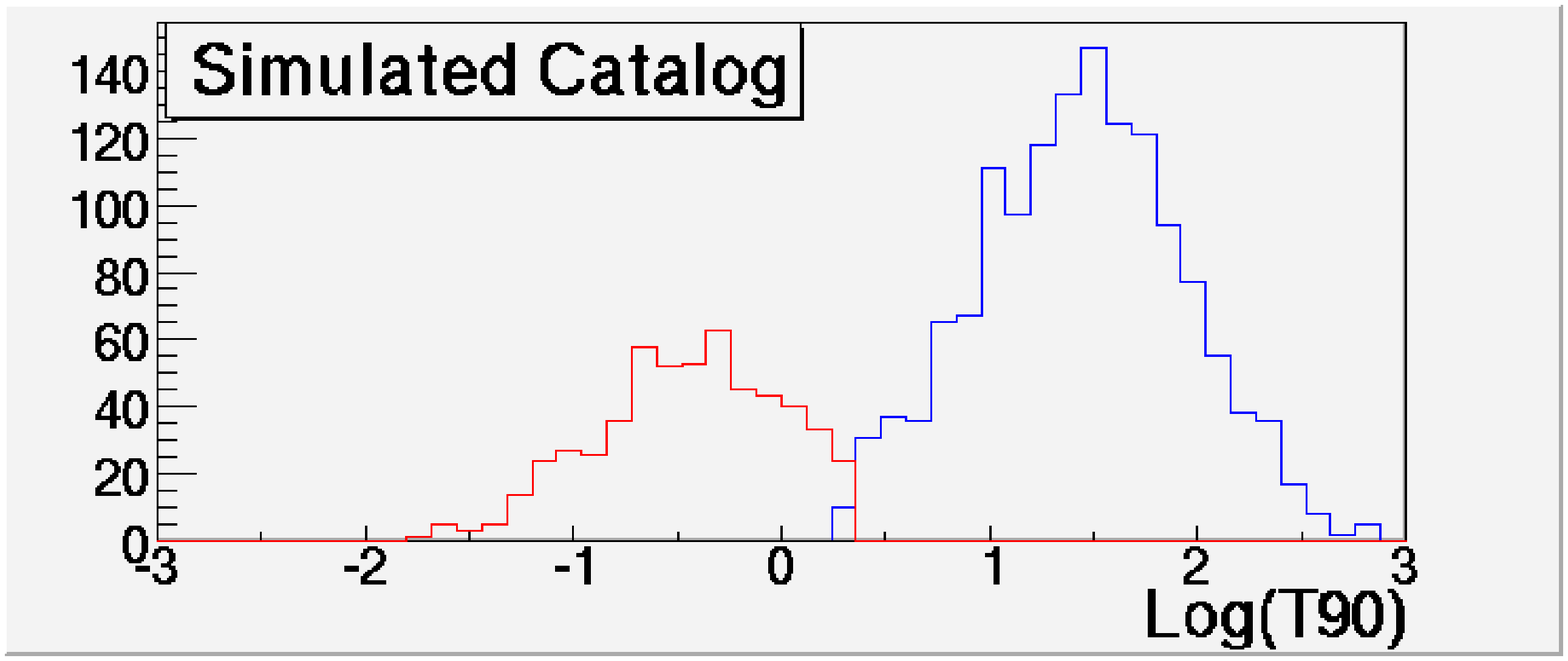}{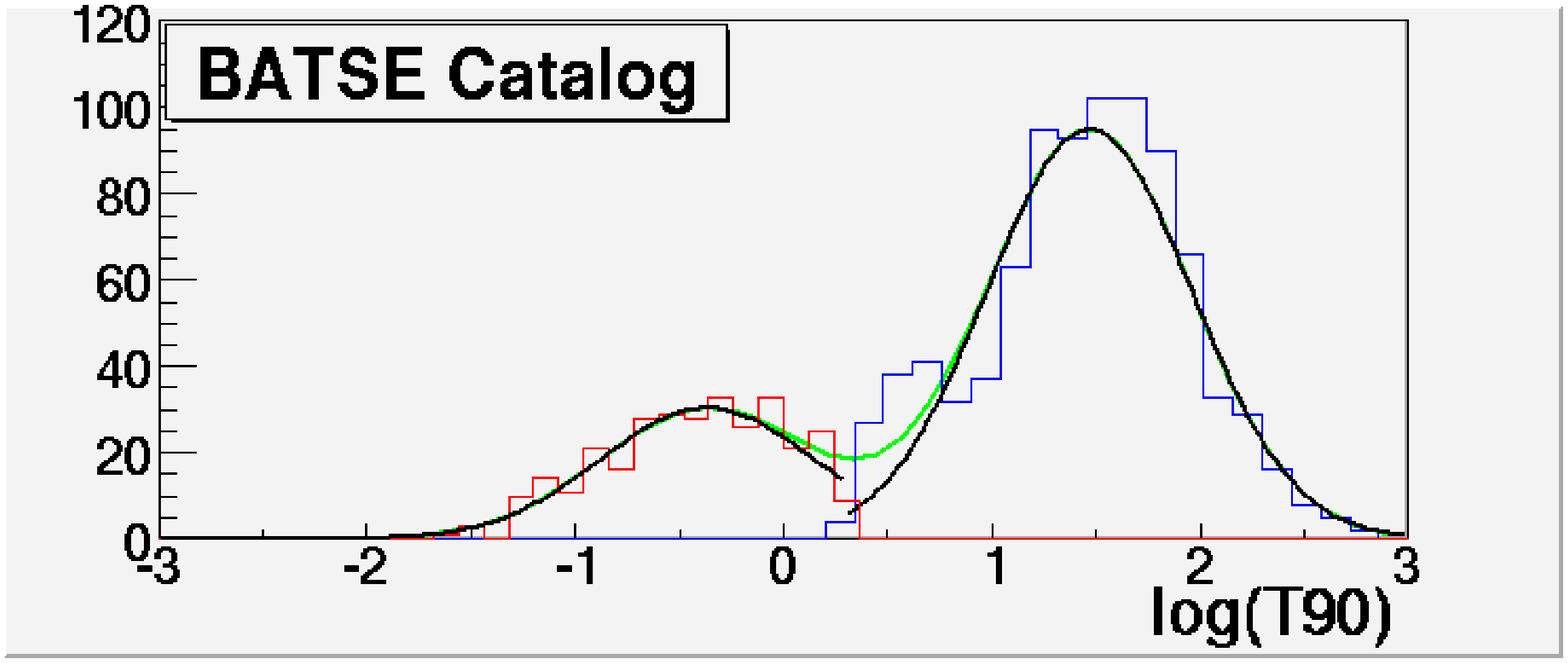}
\caption{Comparison between the observed (left) and the simulated (right) distribution of GRB duration.}
\label{fig1}
\end{figure}

The LAT team has set up a complete simulation chain, including generation of the incoming flux, full simulation of the detector 
resonse, reconstruction chain and analysis of the final trigger and alert. It is to be noted that the flux simulations 
include background (from albedo, cosmic and diffuse gamma ray events) and two possible ways of signal generation: the first based on a 
complete physical model, the second based on a phenomenological extrapolation from BATSE data. Orbit, tilting and correct 
exposure are also taken into account.

The physical model has been validated in the following way: the BATSE observed distribution of duration has been fitted with a double 
gaussian, for short and long bursts (see Figure 1, left panel). Then, we used this bimodal distribution as input parameter 
for our model, which explores the parameter space and returns a burst with a given duration. The resulting simulated catalog is 
depicted in Figure 1, right panel. Finally, the simulated fluences are compared with the observed ones, for the 4 
BATSE channels (Figure 2). In this way we have a calibration method to correlate the emission at high energy with the 
low energy observations. Scaling laws are also reproduced.   

\begin{figure}
\plottwo{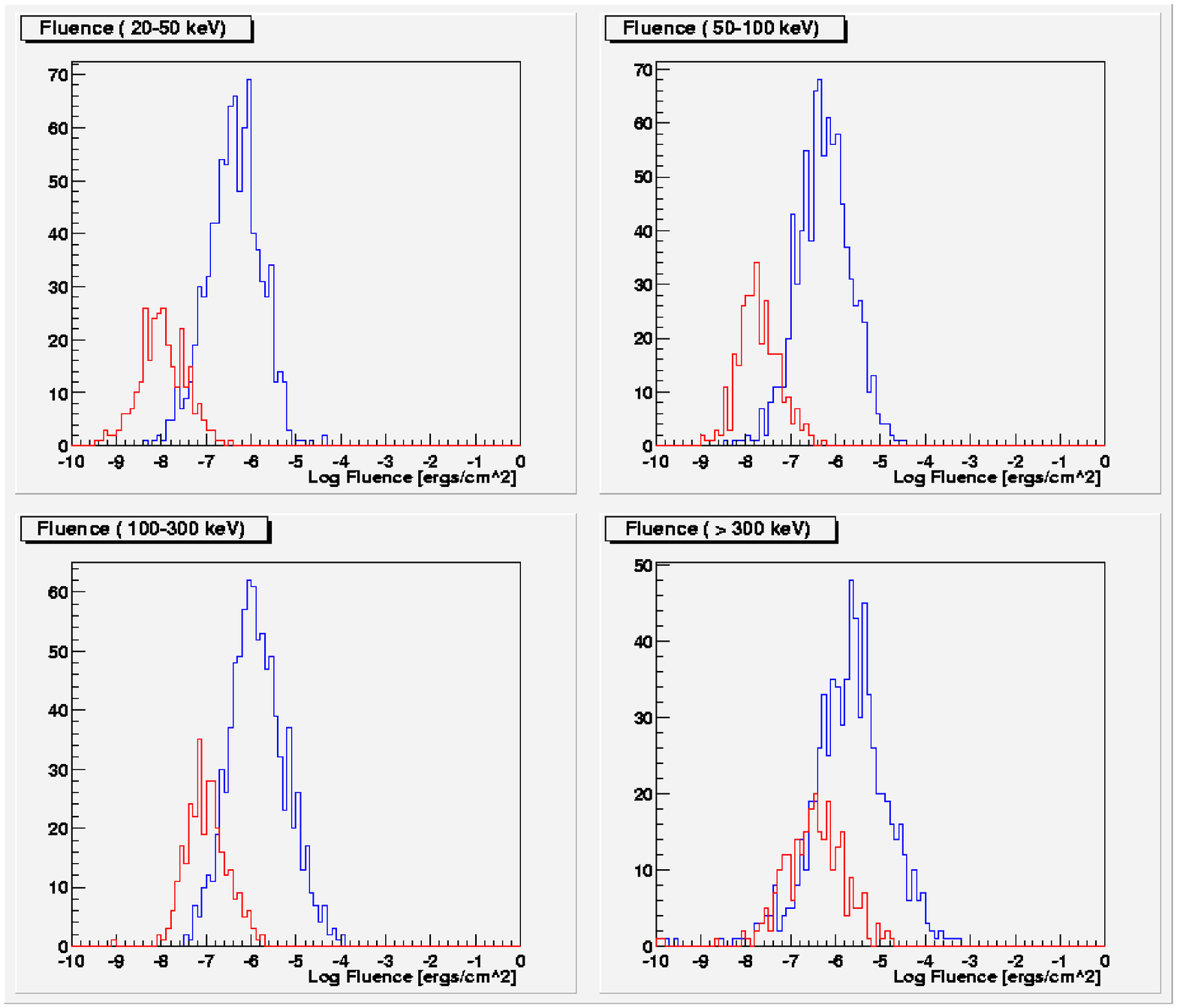}{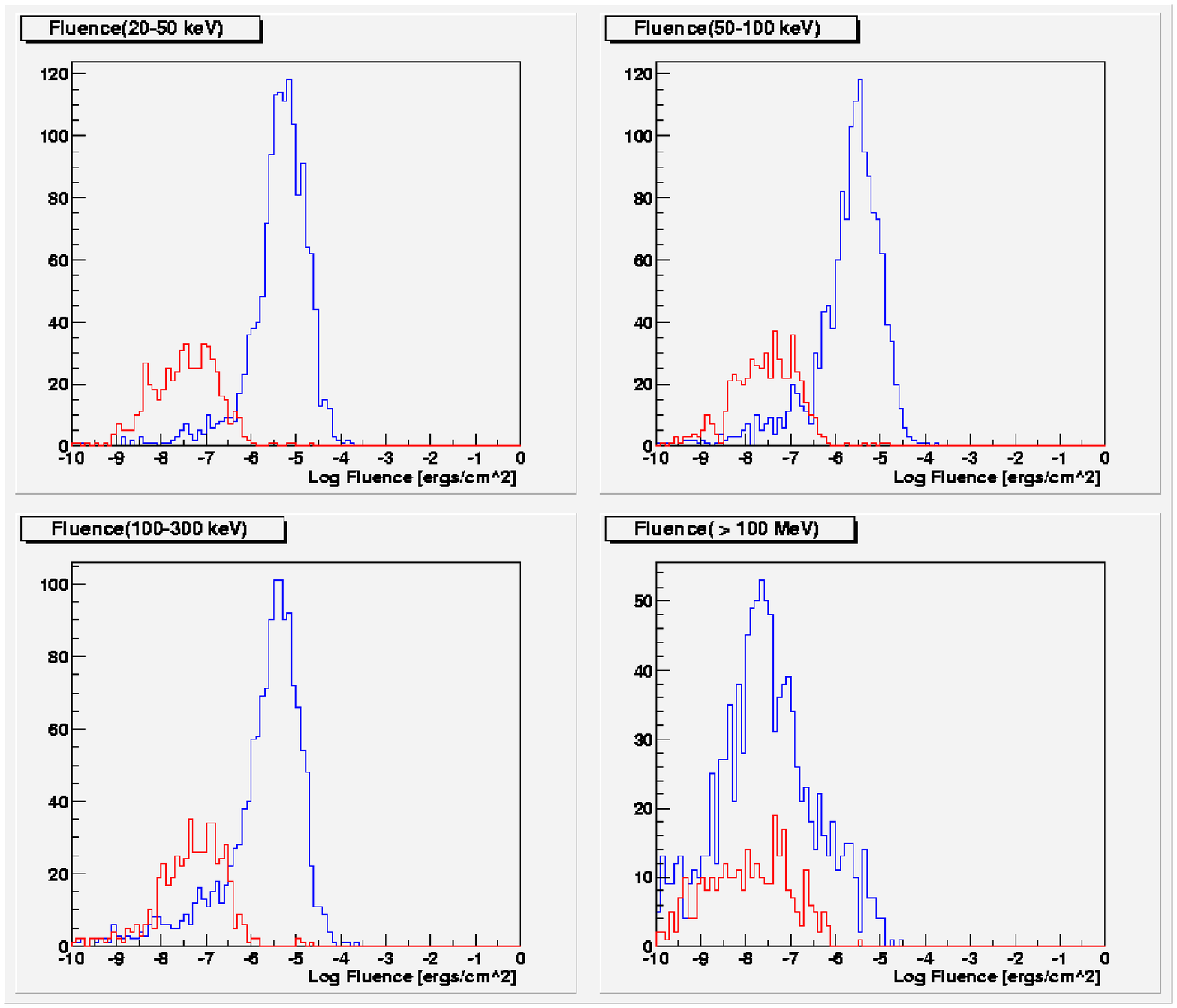}
\caption{Comparison of fluences between the simulated catalog (right) and BATSE observation (left). 
The 4th plot in the right panel shows the estimated emission for short burst (red) and long 
burst (blue) above $100$ MeV.}
\label{fig2}
\end{figure}

\section{Trigger and Alerts}

After calibrating the model, we investigated the efficiency of GLAST in triggering and reconstructing photons coming from GRB events. 
In this preliminary study we have tested various efficiencies for a sample of gamma ray bursts without background. 
A set of GRB's with different fluences above 100 MeV has been generated, and the resulting signal has been processed through the
LAT instrument simulator. Figure 3 (left) shows the number of incident, triggered and reconstructed gammas as a function of 
fluence above 100 MeV. 
The general trend is that for a typical burst of $10^{-8}$ erg/cm$^2$ we expect approximately 100 incident photons, and a photon trigger  
efficiency of ~30\%. More than 90\% of the triggered photons will then be reconstructed. 

\begin{figure}
%\plottwo{cohentanugij1_5.eps}{cohentanugij1_6.eps}
\plotfiddle{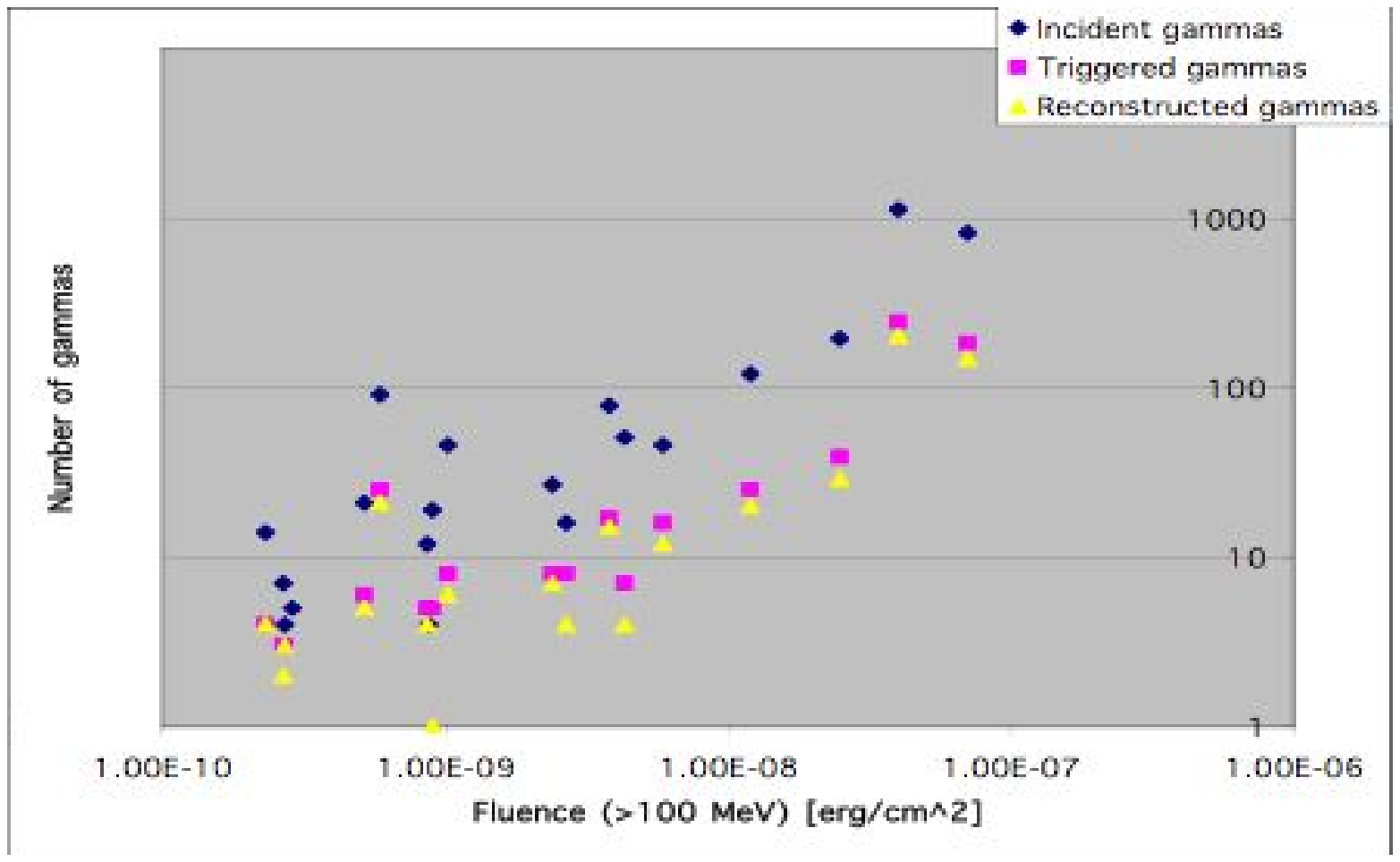}{2.3cm}{0}{40}{60}{-220}{-100}
\plotfiddle{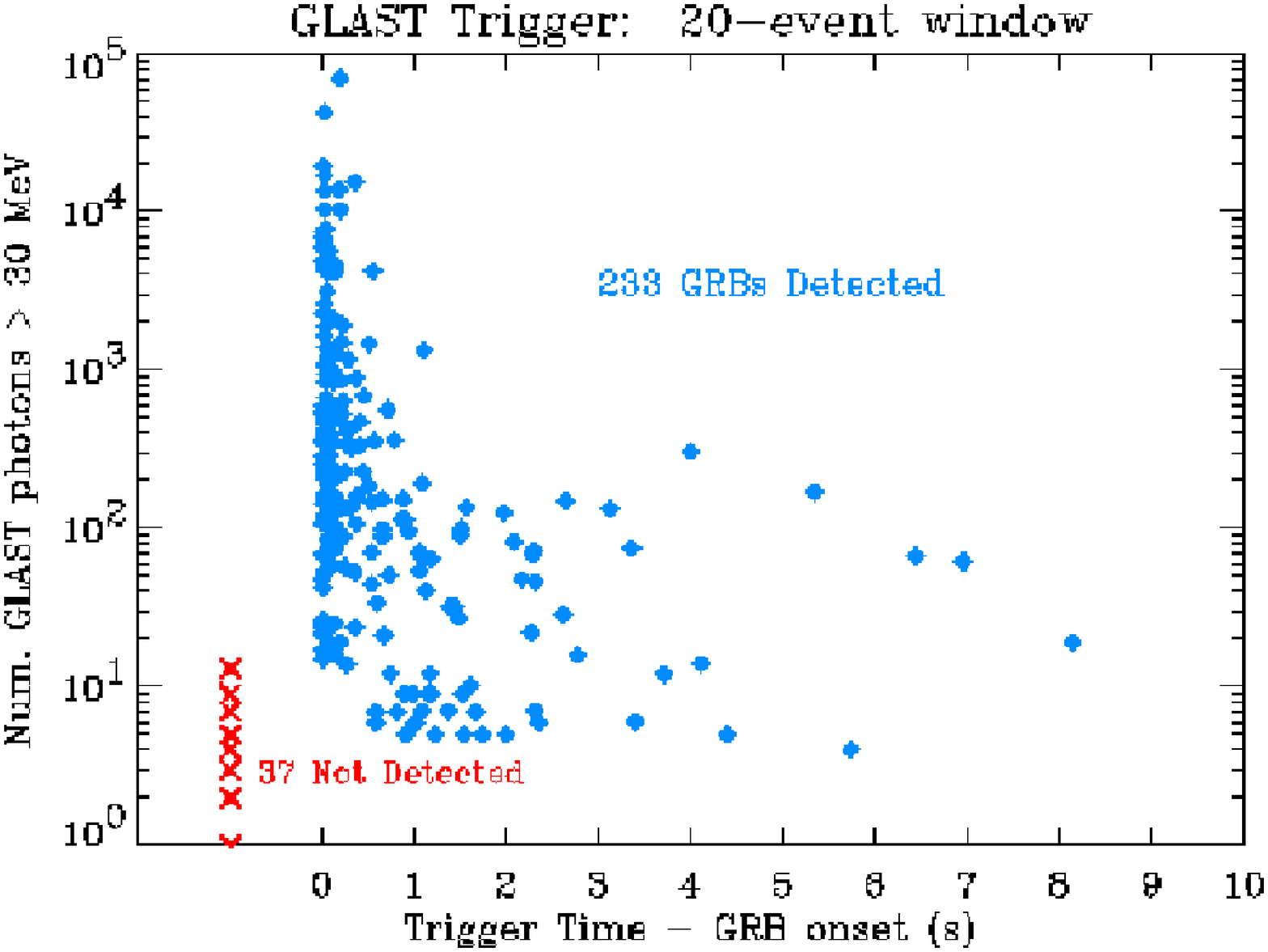}{2.3cm}{0}{30}{30}{-30}{-15}
\caption{Trigger efficiency for several GRBs with different fluences. 
The number of incident, triggered and reconstructed photons are indicated.}
\label{fig3}
\end{figure}

For the purpose of this study, we use a 'strawman' real time unbinned trigger algorithm: the sky is scanned with a sliding 20 
event window, looking for clusters of events in a 35 degree circle; 
then a joint spatial-temporal likelihood is formed. 
The threshold for this trial situation was set adopting an arbitrarily low 
background event rate of 3 Hz, to produce less than one false trigger per week.

The preliminary results of this trigger simulation are presented
in Figure 3.  For the quoted background event rate, efficiency reached
about 85\% for the BASTE-like bursts, with around 78\% of those bursts
detected by GLAST within one second of GRB onset.

\section{Conclusion}
We have investigated the performances of GLAST in detecting GRB events, 
using a complete simulation chain. From our preliminary
results, it appears that GLAST should be able to substantially improve on
the EGRET detection rate of about one GRB per year. In addition to burst
alerts and localizations provided by the GBM, the LAT should
also be able to provide real time triggers and GRB localizations.

Refinements to the trigger simulations will include simulations
at higher background event rates, spatial dependence of diffuse
gamma flux, and temporal dependence of the on board background rates.

\acknowledgements
J.D Scargle acknowledges support from NASA's Applied Information 
Systems Research Program.

\end{document}